\documentstyle[preprint,aps]{revtex}
\def\be{\begin{equation}}
\def\ee{\end{equation}}
\begin{document}

%\begin{center}

\preprint{}
\draft

\title{$Q^2$-dependence of deep inelastic lepton scattering
 off nuclear targets}

\author{O. Benhar $^a$, S. Fantoni $^b$, G.I. Lykasov $^c$
and N.V. Slavin$^c$}
\address{$^a$ INFN, Sezione Sanita', I-00161, Roma, Italy \\
$^b$ Interdisciplinary Laboratory, SISSA,\\
and INFN, Sezione di Trieste, I-34014, Trieste,Italy \\
$^c$ Joint Institute for Nuclear Research,\\
141980 Dubna, Moscow Region, Russia}

\date{\today}

\maketitle

\begin{abstract} 

Deep inelastic scattering of leptons off nuclear targets is analized within the
convolution model
and taking into account nucleon-nucleon correlations. We show that in the
nuclear medium nucleons are distributed
according to a function that exhibits a sizeable $Q^2$-dependence 
 and reduces
to the ordinary light-cone distribution in the Bjorken limit. At 
$Q^2<$~50~(GeV/c)$^2$ and $x>$~1 this $Q^2$-dependence turns out to be 
stronger than the one associated with the nucleon structure function, 
predicted by perturbative quantum chromodynamics.

\end{abstract}


\narrowtext

After the discovery of the European Muon Collaboration (EMC) effect 
\cite{EMC}, a  number of theoretical studies of deep inelastic scattering 
(DIS) off nuclear targets have been carried out within the so called 
convolution model (for recent reviews see \cite{Arneodo,Thomas}).
Within this approach, the nuclear structure function $F_2^A(x,Q^2)$ 
($Q^2 = |{\bf q}|^2 - \nu^2$, where ${\bf q}$ and $\nu$ are the three-momentum 
and energy carried by the exchanged virtual photon, while 
$x=Q^2/2m\nu$, $m$ being the nucleon mass, is the Bjorken scaling variable) is 
written in terms
of the nucleon structure function $F_2^N(x,Q^2)$, that can be extracted from
proton and deuteron data, and the function $f_A(z)$, yielding the 
distribution of the nucleons in the nuclear target as a a function of the
relativistic invariant variable $z$, defined as:
\be
z = \frac{M_A}{m}\frac{(kq)}{(P_Aq)}\ .
\label{def:z}
\ee
In eq.(\ref{def:z}) $P_A\equiv(M_A,0)$, $M_A$ being the target mass, and 
$k\equiv(k_0,{\bf k})$ denote the 
initial four-momenta of the target nucleus and the struck nucleon 
 in the laboratory frame, respectively. More recently, the convolution 
approach has been also extended to study semi-inclusive processes
\cite{Diep,CS,BFLS}

The kinematical region corresponding to DIS is defined by the
Bjorken limit: $Q^2, \nu \rightarrow \infty$ with $x$ finite, 
implying $(|{\bf q}|/\nu) \rightarrow 1$. 
In this limit the quantity $z$ of eq.(\ref{def:z})
can be related to the the light-cone component of the four-momentum 
of the struck nucleon through $z~\rightarrow~k^+/m=(k_0 - k_z)/m$, 
with $k_z=({\bf k}\cdot{\bf q})/|{\bf q}|$.
When the value of $Q^2$ is not large enough and the Bjorken
 limit is not yet reached, however, it is no longer possible to identify
 $z$ with the light-cone variable. As a consequence, in this
regime  
the theoretical nuclear structure function evaluated within the convolution 
model should in principle exhibit a $Q^2$-dependence coming from the 
distribution function $f_A(z,Q^2)$, in addition to the one associated with the 
nucleon structure function $F_2^{N}(x,Q^2)$.

In this paper we report a
calculation of $f_A(z,Q^2)$ in which the kinematical constraints implied
by the Bjorken limits are released, and discuss the relevance of the 
$Q^2$-dependence of the resulting distribution function to the 
interpretation of DIS data.

The distribution function $f_A(z,Q^2)$ is defined as:
\be
f_A(z,Q^2)= z \int d^4k\ S(k)\ 
\delta \left( z - \frac{M_A}{m}\frac{(kq)}{(P_Aq)} \right)\ , 
\label{def:f}
\ee
where
$S(k)$ is the relativistic vertex function. $S(k)$ can be approximated
by the nonrelativistic spectral function $P({\bf k},E)$, yielding 
the distribution in momentum and removal energy of the nucleons in
the target nucleus, according to \cite{BPSEMC}:
\be
S(k)=  \left( \frac{m}{k_0} \right) P({\bf k},E), 
\label{def:S}
\ee
with
\be
k_0 = M_A - \left[ (M_A - m + E)^2 + |{\bf k}|^2 \right]^{1/2}.
\label{def:k0}
\ee
It has to be pointed out that the above defintion of $S(k)$  
can be regarded as a particular implementation of the expansion 
\be
S(k) = P({\bf k},E) \left[ 1 + O \left( \frac{k^2}{m^2}\right) +  
\ldots\  \right]\ ,
\ee
originally proposed in ref.\cite{FSrep}, and that the 
 distribution function obtained using $S(k)$ defined as in eq.(\ref{def:S})
fulfills the normalization requirement
\be
\int dz f_A(z,Q^2) = 1
\ee
for any $Q^2$.
 
Substitution of eq.(\ref{def:S}) into eq.(\ref{def:f}) leads to:
\be
f_A(z,Q^2)=\int_{E_{min}}^{E_{max}} dE \int d^3k 
\left(\frac{m}{k_0}\right)P({\bf k},E)
\delta\left(z-\frac{(kq)}{m\nu}\right),
\label{f}
\ee
where $E_{min}$ denotes the minimum energy required to remove a nucleon from
the target nucleus and $E_{max} = \sqrt{s} - M_A$, with 
$s = (M_A + \nu)^2 - |{\bf q}|^2$.
Equation (\ref{f}) is usually evaluated in the Bjorken limit. 
Here we give a general expression, applicable for any 
$Q^2$ and $\nu$, for the case of infinite nuclear matter, in which 
$f_A(z,Q^2)$ takes a particularly simple and intuitive form. 
The generalization of the nuclear matter result to the case of a
finite size nucleus, in which the recoil of the spectator $(A-1)$-particle
system has to be taken into account, is straightforward.

Integration of eq.(\ref{f}) over cos $\theta = k_z/|{\bf k}|$  
using the delta function yields:
\be
f_A(z,Q^2) = f_A(z,\beta) = \frac{2\pi mz}{\beta}\int_{E_{min}}^{E_{max}}
 dE\int_{k_{min}(E,z,\beta)}^\infty k\ dk \left(\frac{m}{k_0}\right)
P({\bf k},E)\ ,
\label{f:int}
\ee
where
\be
\beta=\frac{|\bf q|}{\nu}=\left( 1+\frac{4m^2x^2}{Q^2} \right)^{1/2}\ ,
\label{def:beta}
\ee
and $k_{min}$ is given by:
\be
k_{min}=\frac{|m(1-z)-E|}{\beta}.
\label{def:kmin}
\ee
Eqs.(\ref{f:int})-(\ref{def:kmin}) show that in the Bjorken limit, 
corresponding to $\beta=(|{\bf q}|/\nu)=1$, the standard expression 
of the light-cone 
distribution is recovered. In general, the distribution function
$f_A(z,\beta)$ depends upon $\beta$ through both the factor $(1/\beta)$ 
in front of the integral of eq.(\ref{f:int}) and through the lower limit 
of the momentum
integration, defined by eq.(\ref{def:kmin}), implying that scattering
processes at different $\beta$ probe the nuclear spectral function at 
different values of $k$ and $E$. This $\beta$-dependence can be readily 
estimated using the approximation employed in refs.\cite{CL,BFLS}, i.e. 
replacing the nucleon removal energy with the average value 
$\langle E \rangle$ in eq.(\ref{def:kmin}):
\be
k_{min}=\frac{|m(1-z)-\langle E \rangle|}{\beta},
\ee
where, using the nuclear matter spectral function of ref.\cite{BFF}, $\langle
E \rangle = 62$~MeV. Within the above approximation, one can easily 
show that the 
inclusion of the $\beta$-dependence produces a shift in $z$, so 
 that $f_A(z,\beta)$ of eq.(\ref{f:int}) can be
 related to $f_A(z,\beta=1)$ calculated at a different value of $z$:
\be
f_A(z,\beta)=\beta \left(\frac{z}{{\widetilde z}}\right)
f_A({\widetilde z},\beta=1)\ ,
\label{shifted:f}
\ee
with
\be
{\widetilde z} = 1 + \frac{(z-1)}{\beta} - \frac{(\beta-1)}{\beta}
\frac{\langle E \rangle }{m} \simeq 1 + \frac{(z-1)}{\beta}
\label{shifted:z}
\ee
The calculated $f_A(z,\beta)$ are bell-shaped, with a maximum around
 $z=1$. According to eq.(\ref{shifted:z}), $\beta > 1$
implies ${\widetilde z} < (>) z$ at $ z > (<) 1$. 
As a consequence, at $\beta > 1$ the peak in $f_A(z,\beta)$ at $z \sim 1$
 gets lower and wider, whereas the tails at $z>1$, where the contributions
 of the high momentum components of the nuclear spectral function dominate, 
 are significantly 
enhanced. These features are clearly illustrated in Fig.1, where 
$f_A(z,\beta)$ resulting from the full calculation (i.e. without
replacing $E$ with $\langle E \rangle$ in eq.(\ref{def:kmin})), 
is plotted as a function of $z$ for different values of $\beta \geq 1$.
Fig. 1 also show that the overall $\beta$-dependence of $f_A(z,\beta)$ is 
sizeable, particularly at $z>1$.

At $z \leq 0.75$ the nuclear matter distribution function $f_A(z,\beta)$ defined by 
eqs.(\ref{f:int})-(\ref {def:kmin}) can be parametrized in the form:
\be
f_A(z,\beta)=A_1{\rm exp}(1-\beta)
{\rm exp}\left[-0.5\left(\frac{z-A_2}{A_3}\right)^2\right]+A_4z^2\ ,
\label{par:1}
\ee
whereas at $z > 0.75$ one can use:
\be
f_A(z,\beta)=zA_5(1+A_7\varphi){\rm exp}(-A_8\varphi)\ ,
\label{par:2}
\ee
where
\be
\varphi=\frac{ | m(1-z)-A_6{\rm exp}[2(1-\beta)] |}{\beta}\ .
\label{phi}
\ee
The numerical values of the parameters appearing in 
eqs.(\ref{par:1})-(\ref{phi}) are: $A_1$=1.74, $A_2$=0.83, $A_3$=0.1, 
$A_4$=0.245, $A_5$=2.28, $A_6$=0.0655, $A_7$=35.6 and  $A_8$=16.6.

The above parametrization of $f_A(z,\beta)$ can be used to study 
 DIS within the convolution model. The nuclear structure function 
$F_2^A(x,Q^2)$ for an isoscalr target is written in the form:
\be
F_2^A(x,Q^2)=\int_x^A dz\ f_A(z,\beta)\ 
F_2^N\left(\frac{x}{z},Q^2\right)
\label{F2A}
\ee
where $F_2^N(x,Q^2)=\left[F_2^p(x,Q^2) + F_2^n(x,Q^2)\right]/2$,  
 $F_2^p(x,Q^2)$ and $F_2^n(x,Q^2)$ being the proton and 
neutron structure functions, respectively, that can be extracted from 
the DIS data off hydrogen and deuterium targets. 

The effect of the $\beta$-dependence associated with $f_A(z,\beta)$
turns out to be very small in the region of the classical EMC effect, 
corresponding to 0.3 $<x<$ 0.8, that has been extensively analyzed using 
the convolution model in the Bjorken limit. On the other hand, 
 at  $x>$1  the $Q^2$-dependence of $F_2^A(x,Q^2)$ 
coming from the distribution function $f_A(z,\beta)$ turns out to
be larger than the one associated with the nucleon structure function 
$F_2^N(x,Q^2)$, predicted by the perturbative quantum chromodynamics (QCD) 
evolution equations \cite{GL,L,AP}
 
In Fig. 2 the large $x$ behaviour of the nuclear matter structure function 
resulting from the approach described in this paper is compared to
the $^{12}$C data taken at CERN by the Bologna-CERN-Dubna-Munich-Saclay (BCDMS) 
collaboration \cite{BCDMS}. The theoretical 
 $F_2^A(x,Q^2)$ has been calculated from eq.(\ref{F2A})
using the parametrized $F_2^N(x,Q^2)$ of ref.\cite{MRRS} and the spectral 
function of ref.\cite{BFF}. Even though the structure function at $x >$ 1 is
mostly sensitive to the short range behaviour of the nuclear wave function, and
is therefore not expected to be strongly affected by finite size effects, the 
comparison 
between infinite nuclear matter and $^{12}$C has to be taken with some 
caution, since the nuclear matter equilibrium density is sizeably higher than 
the average density in the nucleus of $^{12}$C ( 0.16~fm$^{-3}$, compared to 
$\sim$~0.12~fm$^{-3}$). However, since the amount of high momentum components in the nuclear matter wave function decreases as the density decreases, a
 more accurate calculation, carried out using the local density approximation 
 \cite{LDA}, would lower the theoretical curve at large $x$, improving the 
 agreement with the data..

The $Q^2$-dependence of the nuclear matter $F_2^{A}(x,Q^2)$ 
at $x=1.3$ and $x=1.7$ is shown in Fig.3. 
The dashed curves correspond
 to the Bjorken limit, when $\beta=(1+4m^2x^2/Q^2)^{1/2}=1$. In this
case the $Q^2$-dependence of $F_2^A(x,Q^2)$ has to be ascribed
to the nucleon structure function $F_2^N(x,Q^2)$ only.
 The solid curves have been obtained including the $Q^2$-dependence
implied by the $\beta$-dependence of the distribution function $f_A(z,\beta)$, 
parametrized as in eqs.(\ref{par:1})-(\ref{phi}).
 It clearly appears that this dependence produces large effects  
 at $Q^2<50$ (GeV/c)$^2$, particularly at $x=1.7$. 

In conclusion we find that, primarily due to nucleon-nucleon correlations, 
the distribution of 
nucleons in the nuclear medium as a function of the relativistic invariant 
quantity $z$, defined in 
eq.(\ref{def:z}), exhibit an additional dependence upon the ratio 
$x^2/Q^2$, or $Q^2/\nu^2$, in the kinematical regime in which the
Bjorken limit is not yet reached.. The effect of the $\beta$-dependence of 
$f_A(z,\beta)$ on DIS is appreciable at $x>$1, where even small deviations 
of $\beta$ from unity
produce large effects on the nuclear structure function. For example, at
$x=$1.3 and $Q^2$=60 (GeV/c)$^2$, corresponding to $\beta$=1.05, the ratio 
of the nuclear structure functions calculated from eq.(\ref{F2A}) using 
$\beta$=1.05 and $\beta$=1, respectively, is $\sim$ 1.7. Fig. 3 also shows that
the effect becomes larger as $x$ increases. It has to be emphasized that
at $x>$1 the $Q^2$-dependence discussed in the present work results in a 
$Q^2$-dependence of the nuclear structure function $F_2^A(x,Q^2)$ much stronger 
than the one associated with the nucleon structure function, particularly 
at $Q^2<$50 (GeV/c)$^2$. Unfortunately, the only DIS data available at 
$x>$1 cover the region $x<$1.3 and $Q^2>$ 60 (GeV/c)$^2$ and only provide
upper bounds of the nuclear structure function. However, the effect discussed
in the present work will certainly be critical in the analysis of the
planned DIS experiment at $x>$1 and moderate $Q^2$ \cite{ELFE}. 

\acknowledgments

This work has been encouraged and supported by the Russian Foundation
of Fundamental Research. We gratefully acknowledge very helpful
discussions with A.Fabrocini, E.Oset and E.Marco. One of us (GIL) wishes
to thank S. Fantoni for the kind hospitality at the Interdisciplinary
Laboratory of SISSA, where part of this work has been carried out.

\begin{figure}
\caption{ $z$-dependence of the distribution function $f_A(z,\beta)$, 
calculated for infinite nuclear matter at equilibrimu density, at 
different values
of $\beta$. Solid line: $\beta=1.0$; dashed line: $\beta=1.1$; dot-dash
line: $\beta=1.2$.} 
\end{figure}

\begin{figure}
\caption{ $F_2^A(x,Q^2)$ for infinite nu\-cle\-ar mat\-ter at 
e\-qui\-li\-brium density, 
calculated using eqs.(8)-(10) and (17), (solid lines), compared to the carbon 
structure function measured by the BCDMS collaboration [14].}
\end{figure}

\begin{figure}
\caption{$Q^2$-dependence of the nuclear matter $F_2^A(x,Q^2)$ at different 
values of $x$. The dashed lines correspond to the Bjorken limit, i.e. to
$\beta = 1$, whereas the solid lines have been obtained including the
$Q^2$-dependence of the distribution function $f_A(z,\beta)$, 
parametrized as in eqs.(14)-(16). The experimental upper bounds are from
ref. [14].}
\end{figure}

% \newpage
% \begin{center}
% \mbox{\epsfig{file=omfig1.ps,width=\textwidth}}
% \end{center}
\end{document}